\documentclass[twocolumn,pre,floatfix]{revtex4}
\usepackage{psfrag,epsfig,amsfonts,amssymb,amsmath,wasysym,bm}
\usepackage{dcolumn}
\usepackage{bbold}
\usepackage[normalem]{ulem}
\usepackage{color}
\usepackage{tabularx}

\usepackage{enumitem} % enumerated lists

\usepackage{hyperref}
\newcommand{\ket}[1]{\lvert #1 \rangle} 	% ket
\newcommand{\bra}[1]{\langle #1 \rvert}	% bra
\newcommand{\ketN}[1]{\lvert #1 \rangle_{\! 0}} 	% ket
\newcommand{\braN}[1]{{_0\!}\langle #1 \rvert}	% bra
\newcommand{\e}{\mathrm{e}}		% Euler number
\newcommand{\I}{\mathrm{i}}		% imaginary unit
\newcommand{\lvsp}{\varepsilon}

\newcommand{\rhomc}{\rho_{\mathrm{mc}}}

\newcommand{\tr}{\mbox{Tr}}
\newcommand{\At}{{\cal A}_t}
\newcommand{\Abar}{\overline{{\cal A}_t}\,}

\providecommand{\av}[1]{[\,#1]_V}

\newcommand{\lmat}{\left( \begin{matrix}}	% matrix begin
\newcommand{\rmat}{\end{matrix} \right)}	% matrix end
\definecolor{revcolor}{RGB}{32,64,255}

\begin{document}

\title{Refining Deutsch's approach to thermalization}
%\title{Complementing \LD{Refining?} Deutsch's approach to thermalization}
%\title{Status report on Deutsch's approach to thermalization}
%\title{Implications of Deutsch's approach to thermalization}
%\title{Typicality of thermalization}
\author{Peter Reimann}
%\email{reimann@physik.uni-bielefeld.de}
\author{Lennart Dabelow}
%\email{ldabelow@physik.uni-bielefeld.de}
\affiliation{Fakult\"at f\"ur Physik, 
Universit\"at Bielefeld, 
33615 Bielefeld, Germany}
\date{\today}

\begin{abstract}
The ground breaking investigation by Deutsch [Phys.~Rev.~A 43, 2046 (1991)]
of how closed many-body quantum systems approach
%on the relaxation of isolated many-body systems towards
thermal equilibrium is 
revisited.
%further developed 
%\PR{It is pointed out that some important steps were still missing in that paper and it is  shown how those gaps can be closed.}
It is shown how to carry out some important steps which were still missing in that paper.
Moreover, the class of admitted systems is considerably extended.
%\LD{Moreover, the results are extended to a larger class of setups.}
\end{abstract}

\maketitle

%%%%%%%%%%%%%%%%%%%%%%%%%%%%%%%%%%%%%%%%%%%%%%%%%%%%%%%

The very common empirical observation that closed macroscopic 
systems approach thermal equilibrium after sufficiently long times
is still very hard to deduce 
directly from a basic quantum mechanical description 
\cite{tas16,dal16,gog16,mor18,bor16}.
The seminal 1991 paper by Deutsch \cite{deu91}
is nowadays mostly associated with the so-called
eigenstate thermalization hypothesis (ETH) 
%issue 
\cite{dal16,gog16,mor18},
but its actual main focus was in fact on the 
above-mentioned issue to better understand thermalization 
on a microscopic basis.
The objective of our present 
paper is to complement some essential steps along 
these lines which were
% still missing
anticipated, but not derived stringently 
in Deutsch's original work \cite{deu91}.
%Note is to complement some of the steps along these lines which were still 
%missing in Deutsch's original work \cite{deu91}.
%Since many of the essential ingredients are known, 
%we omit most of the technical details, yet it seemed
%to us worthwhile to collect here those widely scattered 
%fragments into a coherent account.

As usual \cite{gog16,dal16,bor16,mor18,deu91},
we consider a system with Hamiltonian $H$
and concomitant eigenvalues $E_n$ and 
eigenstates $\ket{n}$, which is prepared in a 
pure or mixed initial state $\rho(0)$ and
then evolves according to 
$\rho(t) = \e^{-\I H t} \, \rho(0) \, \e^{\I H t}$
($\hbar = 1$).
The expectation value of an observable 
(self-adjoint operator) $A$ thus takes 
the form
\begin{eqnarray}
        \At 	
%        \< A \>_{\! \rho(t)} 
	:= \tr\{\rho(t) A\}
	= \sum_{m, n} \e^{\I (E_n - E_m) t} \, \rho_{mn}(0) \, A_{nm} \,,
\label{1}
\end{eqnarray}
where $\rho_{mn}(t) := \bra{m} \rho(t) \ket{n}$,
$A_{nm} := \bra{n} A \ket{m}$,
and the summation is over all states.
In case of degeneracies, the corresponding
eigenstates $\ket{n}$ are chosen so that
the matrix $\rho_{mn}(0)$ is diagonal within 
every eigenspace.
Indicating averages over all times $t\geq 0$ by an overline,
one readily can conclude that
%Denoting time-averages by an overline, it follows that
\begin{eqnarray}
%\Abar := 
\overline{\At}
%	\overline{ {\< A \>_{\! \rho(t)}} }
= \sum_{n} \rho_{nn}(0) \, A_{nn} 
	\ .
\label{2}
\end{eqnarray}

The first essential point of Deutsch's approach \cite{deu91} 
is to focus on Hamiltonians of the form
\begin{eqnarray}
H = H_0 +V 
\ ,
\label{3}
\end{eqnarray}
consisting of an unperturbed $H_0$ with eigenvalues 
$E^0_\nu$ and eigenstates $\ketN{\nu}$, and
a sufficiently weak perturbation $V$ with matrix 
elements $V^0_{\mu\nu} := \braN{\mu} V \ketN{\nu}$.
%\LD{In the original setup from Ref.~\cite{deu91}, $H_0$ is introduced to describe a system of noninteracting particles and $V$ their interactions, but other splittings are also conceivable.}
For instance, $H_0$ may describe a 
system of noninteracting particles 
and $V$ their interactions \cite{deu91}, 
but various other examples are also conceivable
\cite{f0}.
Instead of working with the perturbation $V$
pertaining to one specific system of interest, 
the key idea of Ref. \cite{deu91} is to approximately 
model that perturbation by a random matrix.
More precisely, the $V^0_{\mu\nu}$
are considered to be sampled from a suitably
chosen random matrix ensemble, which
ideally should still capture the main characteristics 
of some realistic interaction sufficiently well.

Such a randomization of $V$ entails
a corresponding randomization of the
eigenstates $\ket{n}$ of $H$ via (\ref{3})
and hence of the matrix elements on the
right-hand side of (\ref{2}).
But since the initial condition $\rho(0)$ and the 
observable $A$ themselves are {\em not}
randomized, it is convenient to employ the
overlaps 
\begin{equation}
	U_{m\nu} := \langle m | \nu \rangle_{\!0}
\label{4}
\end{equation}
between perturbed and unperturbed eigenstates, 
and then to rewrite (\ref{2}) in the form
\begin{eqnarray}
\Abar
%	\overline{ \< A \>_{\!\rho(t)} }
        & = & 
	\sum_{\mu_1, \mu_2, \nu_1, \nu_2} \rho^0_{\mu_1 \nu_2}(0) \, A^0_{\mu_2 \nu_1} 
	W^{\mu_1 \mu_2}_{\nu_1 \nu_2}
	\,,
\label{5}
\\
	W^{\mu_1 \mu_2}_{\nu_1 \nu_2}
        & := & \sum_{n} 
          U_{n \mu_1} U_{n \mu_2} 
          U^*_{n \nu_1} U^*_{n \nu_2} \ ,
\label{6}
\end{eqnarray}
where $\rho^0_{\mu\nu}(0) := \braN{\mu} \rho(0) \ketN{\nu}$
and $A^0_{\mu\nu} := \braN{\mu} A \ketN{\nu}$.
The entire statistics of the random quantity (\ref{5}) is
thus encapsulated in the last factor on the right-hand side,
while the first two factors are kept fixed (non-random).

Deducing the statistical properties of the random variables
$W^{\mu_1 \mu_2}_{\nu_1 \nu_2}$ in (\ref{6}) from those 
of the random matrices $V^0_{\mu\nu}$
is a technically very demanding task.
Accordingly, analytical progress is up to 
now only possible for certain classes of 
random matrices $V^0_{\mu\nu}$.
A relatively simple class has been considered by
Deutsch, and will be specified in the next paragraph,
while more general classes will be addressed later on.
For all those random matrix ensembles,
%which will be considered later on, 
it will turn out that a particularly important role is 
played by the second moments 
$\av{\lvert U_{m\nu} \vert^2}$ of the 
overlaps from (\ref{4}),
where the symbol $\av{\,\cdots}$ indicates 
the ensemble average.
Moreover, those second moments
$\av{\lvert U_{m\nu} \vert^2}$ will turn out not to 
depend separately on $m$ and $\nu$, but only on 
the differences $m-\nu$,
i.e., the function
%\LD{[Die Anmerkung, dass eine konstante Zustandsdichte 
%angenommen wird (siehe unterhalb von~\eqref{9}) sollte vielleicht hier schon hin.
%\PR{Auch Annahmen \"uber die $V^0_{\mu\nu}$ sind n\"otig ...;
%all das w\"urde den nat\"urlichen Fluss hier zu sehr unterbrechen,
%daher ``... will turn out ...''.}]}
%%there exists a function $u(n)$ with the property that \cite{f1}
\begin{equation}
u(m-\nu) := \av{\lvert U_{m\nu} \vert^2} 
%	\av{\lvert U_{m\nu} \vert^2} = u(m-\nu) 
%\, .
\label{7}
\end{equation}
is well-defined \cite{f1}.

Following Deutsch \cite{deu91}, let us first
focus on real symmetric $V$-matrices, whose matrix 
elements $V^0_{\mu\nu}$ are identically
distributed, independent (apart from $V^0_{\mu\nu}=V^0_{\nu\mu}$)
Gaussian random variables of 
mean zero and variance $\sigma^2$.
In other words, $V^0_{\mu\nu}$ is drawn from the Gaussian Orthogonal Ensemble (GOE).
The first main result of Ref.~\cite{deu91} 
can then be written in the form \cite{f2}
\begin{eqnarray}
u(n) & = & \frac{1}{\pi}\frac{w}{w^2+n^2}
\ ,
\label{8}
\\
w & := & \pi\sigma^2/\lvsp^2
\ ,
\label{9}
\end{eqnarray}
where $\lvsp$ is the mean level spacing of the unperturbed 
energies $E^0_\nu$.
In doing so, it is tacitly taken for granted 
that
%such
an approximation in terms of a 
constant mean level spacing is justified by focusing 
on initial states $\rho(0)$ with a sufficiently 
narrow energy distribution 
%for the unperturbed as well as the perturbed systems 
with respect to both the perturbed and the 
unperturbed systems
in (\ref{3}). 

%We furthermore note that, according to 
In view of (\ref{4}) and (\ref{7}), the parameter $w$
in (\ref{8}) essentially quantifies the perturbation strength 
in terms of how many of the unperturbed
eigenstates notably contribute to the perturbed ones. 
%due to the interaction $V$ in (\ref{3}).
Accordingly, $w\lvsp$ is the corresponding energy range,
over which the unperturbed energy levels are effectively
coupled via the perturbation. 
Generically, the mean level spacing $\lvsp$ is
exponentially small in the system's degrees of freedom \cite{deu91},
hence it is intuitively quite clear (see also \cite{dab20})
that the perturbation will not have any notable effect 
on the unperturbed relaxation unless $w$ 
is very large (exponentially large in the 
degrees of freedom).
% \cite{dab20}). 
From now on we thus always take for granted that
\begin{eqnarray}
w \gg 1
\ .
\label{10}
\end{eqnarray}
In fact, the same assumption is already required 
in Ref. \cite{deu91}, and Eq. (\ref{8}) actually 
amounts to a leading order approximation in the 
small parameter $1/w$.

On the other hand, the perturbation $V$ is also required 
to remain sufficiently weak in the following sense \cite{deu91}:
If both the perturbed and the unperturbed systems {\em were} 
at thermal equilibrium (with the same energy expectation
values as for the ``true'', 
%and in general far from equilibrium
generally far-from-equilibrium
initial state $\rho(0)$), then the corresponding thermal
equilibrium properties (temperature, heat capacity, 
expectation value of $A$ etc.)
%would
should
be approximately 
the same for both systems.
Technically speaking, this requirement
is due to the assumption 
below (\ref{9}) that also the perturbed system must
exhibit a narrow energy distribution \cite{deu91,dab20}.

In a next step (see unpublished item in Ref. \cite{deu91}),
it is assumed that the overlaps (\ref{4}) can be approximately 
considered as Gaussian distributed and statistically 
independent of each other. 
%Hence,
In this case,
the ensemble average in (\ref{6})
can be readily expressed by means of the Isserlis
(or Wick) theorem in terms of the second moments from (\ref{7}),
yielding for the ensemble-averaged time average in 
(\ref{5}) the final main result 
\cite{deu91}
\begin{eqnarray}
\av{\Abar}
        & = & 
	\sum_{\mu\nu} \rho^0_{\mu\mu}(0)\, \tilde u(\mu-\nu)\, A^0_{\nu\nu} 
%	=\sum_{\nu} \tilde \rho^0_{\nu\nu}\, A^0_{\nu\nu} 
	\ ,
\label{11}
\end{eqnarray}
where $\tilde u(n)$ is the convolution of $u(n)$ 
from (\ref{7}) with itself,
\begin{eqnarray}
\tilde u(n) :=\sum_m u(n-m)\,u(m)
\ .
\label{12}
\end{eqnarray}
Introducing the auxiliary density operator
\begin{eqnarray}
\tilde\rho:=\sum_\nu \tilde\rho^0_{\nu\nu} \,\ketN{\nu}\braN{\nu}
\ , 
\label{13}
\end{eqnarray}
which is diagonal in the unperturbed basis with diagonal matrix elements
\begin{eqnarray}
\tilde \rho^0_{\nu\nu} :=
\sum_{\mu} \rho^0_{\mu\mu}(0)\, \tilde u(\mu-\nu)
\ ,
\label{14}
\end{eqnarray}
one can readily rewrite (\ref{11}) as
\begin{eqnarray}
\av{\Abar} =\sum_{\nu} \tilde \rho^0_{\nu\nu}\, A^0_{\nu\nu} = \tr\{\tilde\rho A\}
	\ .
\label{15}
\end{eqnarray}

Taking for granted that $\rho(0)$ exhibits a sufficiently narrow energy 
distribution and the perturbation 
%strengths $w$ in (\ref{9}) is not too large 
is not too strong
(see discussion below Eq. (\ref{9})), 
one can infer from (\ref{8}), (\ref{12}), and (\ref{14}) 
that the energy distribution of $\tilde \rho$ will
also
%also the energy distribution of $\tilde \rho$ will still 
be quite narrow and, at the same time, the level populations 
$\tilde \rho^0_{\nu\nu}$ will vary relatively smoothly 
as a function of $\nu$ due to (\ref{10}).
In other words, $\tilde\rho$ 
%closely resembles 
is rather similar to
the pertinent microcanonical ensemble $\rhomc$ which 
describes the thermal equilibrium properties of the 
unperturbed many-body system in (\ref{3}).
In particular,
it seems reasonable to conjecture that
the expectation value on the right-hand side
of (\ref{15}) will be close to the unperturbed thermal
equilibrium value 
%\cite{f3}
\begin{eqnarray}
\tr\{\tilde\rho A\}\simeq \tr\{\rhomc A\}
\ .
\label{16}
\end{eqnarray}
Notably, the above line of reasoning does not require
%Notably, it is not assumed here 
that the {\em unperturbed} 
system with initial state $\rho(0)$
approaches thermal equilibrium in the long run.
%\PR{does thermalize} \LD{thermalizes?}. 
%exhibits thermalization.
%Rather, the right-hand side of (\ref{16})
%refers to the behavior 
%\PR{if the unperturbed system {\em were}}
%in case the unperturbed system were 
%right from the beginning in
The density operator $\rhomc$ on the right-hand side of~(\ref{16}) merely 
corresponds to the microcanonical ensemble associated
%the microcanonical 
%state $\rhomc$
with the same (macroscopic) energy as 
the true initial state 
$\rho(0)$, i.e., $\tr\{\rhomc H_0\}=\tr\{\rho(0) H_0\}$,
regardless of the actual unperturbed dynamics.

Exploiting that the expectation values at thermal 
equilibrium of the perturbed and unperturbed 
systems can be considered as approximately 
equal (see below (\ref{10})), we finally can conclude 
with (\ref{15}) and (\ref{16}) that the perturbed systems 
exhibit (approximately) thermal expectation values on 
the average over all times $t \geq 0$ and over all members 
$V$ of the considered random matrix ensemble.

As far as the issue of thermalization is concerned 
(see introduction), this is the main finding of Deutsch's 
seminal work \cite{deu91}. While it is clearly a highly non-trivial and
interesting achievement in itself, there still remain 
essentially three unsolved problems. 
(i) 
For many (or even
most) members $V$ of the considered ensemble, the value
of $\Abar$ could in principle still be quite different from the
ensemble average in (\ref{15}), hence it would not be right to
consider such systems as exhibiting thermalization. 
(ii) 
Likewise, the time-dependent expectation values $ \At$ in
(1) could still maintain non-negligible fluctuations about
their time-average $\Abar$ {\em ad infinitum}, and therefore not
exhibit thermalization in any meaningful sense. 
(iii) 
Finally, also the justification of the approximation (\ref{16}) in
the text above Eq. (\ref{16}) may be regarded as not entirely
satisfying: Even though the energy distribution of $\tilde\rho$ 
is narrow and the level populations $\tilde \rho^0_{\nu\nu}$
vary slowly as a function of $\nu$ (see above Eq. (\ref{16})), 
the factors $A^0_{\nu\nu}$ in (\ref{15}) may in principle still 
conspire in such a way that (\ref{16}) is violated.

To settle these issues (i)-(iii) is the main objective of our 
present paper. Before doing so, also some 
remarks, extensions, and precursors 
regarding Deutsch's approach itself are worth mentioning.
%Finally, the expectation values at thermal equilibrium 
%of the perturbed and unperturbed systems can be 
%considered as approximately equal as discussed below (\ref{10}).
%Altogether, according to (\ref{15}) and (\ref{16})
%the perturbed systems are thus found to 
%thermalize on the average over all times $t\geq 0$
%and over all members $V$ of the considered random
%matrix ensemble
%%\LD{even if the unperturbed system fails to thermalize [steht im 
%%Prinzip auch schon in der Fu\ss note \cite{f3}, aber schadet 
%%meines Erachtens nicht, das nochmal ``prominenter'' festzuhalten.
%%\PR{Bin dagegen: Das ist das zentrale Statement, und sollte daher m\"oglichst
%%pr\"agnant sein, d.h. ohne alle nicht unberdingt erforderlichen``Zus\"atze''.
%%Kommt ja auch sp\"ater nochmals ...}]}
%\cite{deu91}.
%
%The above main finding of Ref. \cite{deu91} is clearly a very
%important step into the right direction, but equally clearly, 
%further steps are needed. 

Our first remark is that 
a similar result as in (\ref{8})       
%an even somewhat more general
%\LD{[wei\ss\ nicht, ob Wigner ``allgmeiner'' ist; die Bandstruktur 
%gibt es bei Deutsch nat\"urlich nicht, daf\"ur sind bei Wigner alle 
%Matrixelemente innerhalb des Bandes $\pm v_0$ mit zuf\"alligen Vorzeichen]}
%result than in (\ref{8}) 
%(for somewhat different ensembles of perturbations)    
has actually been obtained 
already  in 1955 by Wigner \cite{wig55}.
Furthermore, the derivations of those results
%of (\ref{8}) 
both by Deutsch and by Wigner are based on 
%non-rigorous
some uncontrolled
assumptions and approximations.
%A %fully rigorous 
%well-controlled 
%derivation (for asymptotically large $w$)
%by means of supersymmetry
%techniques is due to Fyodorov et al. \cite{fyo96}.
An asymptotically exact derivation for large $w$
by means of supersymmetry
techniques has been elaborated 
by Fyodorov et al. in Ref. \cite{fyo96}.
Moreover, they also admitted considerably more 
general ensembles of real symmetric random 
matrices $V^0_{\mu\nu}$,
allowing largely arbitrary (non-Gaussian) distributions for 
the $V^0_{\mu\nu}$ and explicitly including sparse
% in particular sparse (and thus 
%non-Gaussian distributed)
and/or banded random matrices.
In Ref.~\cite{dab20}, the same approach was generalized 
to complex Hermitian random matrices.
Further extensions, e.g., to cases with particularly
strongly fluctuating diagonal random matrix 
elements can be found in Ref. \cite{fyo95}.
Remarkably enough, in all these cases, the Breit-Wigner
distribution from Eq. (\ref{8}) remains a very good 
approximation, at least as long as 
the perturbation strength $w$ in (\ref{9}) is much
smaller than the band width (in case of banded random matrices),
and provided $\sigma^2$ is defined as the variance
$\av{|V^0_{\mu\nu}|^2}$ of the (possibly complex)
random matrix elements $V^0_{\mu\nu}$ close
to the diagonal (i.e., ``locally averaged'' over many
$\mu$ and $\nu$ with relatively small but 
non-vanishing differences $\mu-\nu$).
More precisely speaking, 
it was pointed out already in Refs. \cite{wig55,deu91}
and elaborated more rigorously in
Ref. \cite{fyo96} that a finite band width 
%inevitably must give 
gives
rise to a crossover
of the relatively slow decay of (\ref{8}) 
with $n$ into a much faster (essentially 
exponential) decay for sufficiently 
large $n$, and similarly for
$\tilde u(n)$ in (\ref{12}).
Since physically reasonable perturbation matrices exhibit a finite energy range,
%it is reasonable to expect that
this scenario applies very generally
to
%physically
realistic models \cite{deu91}.
Possible concerns regarding the slowly 
decaying tails of $\tilde u(n)$ in 
Eqs. (\ref{11}), (\ref{14}), and (\ref{15})
can thus be readily 
overcome.
When $w$ is no longer small compared to the 
band width of $V^0_{\mu\nu}$,
%For even stronger perturbations 
%($w$ not small compared to the matrix's band width), 
the second moments from (\ref{7}) 
in general deviate from the 
Breit-Wigner form (\ref{8}), but all the more 
general conclusions turn out to remain 
essentially unchanged. % \PR{\cite{f3}}.
For the sake of simplicity, we thus tacitly focus
from now on cases where (\ref{8}) applies.

As mentioned above Eq. (\ref{11}), an important 
ingredient of Deutsch's approach is the 
assumption that the overlaps in (\ref{4}) are 
(approximately) Gaussian and independent.
However, it was observed in Ref. \cite{nat18}
(see also \cite{dab20})
that this assumption may actually lead to provably 
wrong conclusions in some particular 
(but still relevant) cases.
Therefore, Deutsch's 
%non-rigorous methods were
approximative methodology was
generalized in \cite{nat18} to directly
determine the ensemble average over
products of up to four $U$-matrix elements, 
as appearing on the right-hand side 
of (\ref{6}).
While the so-obtained results now 
indeed exhibit correlations between different 
$U$-matrix elements \cite{nat18}, the  
final relation in (\ref{15}) turns out
to remain unchanged.
However, those approximations from \cite{nat18}
still violate some basic orthonormality properties 
of the unitary matrix $U_{m\nu}$
(see equation (48) in \cite{nat18}).
Apparently, this problem is due to the
approximative character of the employed 
methodology, and it indicates that 
%the implications for the ensemble 
%averages required in (\ref{6}) 
also the resulting prediction (\ref{15}) 
can still not be considered as entirely 
convincing from a more rigorous 
viewpoint.
However, a satisfying resolution of the problem
can be achieved by way of generalizing the 
supersymmetry approach from Ref.~\cite{fyo96} along the lines of
Ref.~\cite{dab20}.
Referring to the Supplemental Material of the latter
publication for the
% rather involved methodology, 
technical details,
one recovers as the final outcome 
of such a lengthy and
% complicated
involved calculation
once again the same result as in (\ref{15}).
Though this result (\ref{15}) thus appears to
be remarkably robust against not very well controlled 
approximations, such a confirmation
by more stringent methods seems noteworthy to us.
%Finally, the problem has been solved in
%Ref. \cite{dab20} by way of generalizing the 
%supersymmetry techniques from Ref. \cite{fyo96},
%%\LD{(see also \cite{ith18} for a different approach via Lippmann-Schwinger-type equations)}, 
%yielding once again the same final result as in (\ref{15}).
%%\LD{[Hier schon Anmerkung, dass jetzt auch f\"ur das ETH-Argument 
%von Deutsch, $A_{nn} \simeq \av{ A_{nn} } = \sum_\mu A^0_{\mu\mu} \, u(n - \mu)$, 
%die Varianz $\av{ (A_{nn})^2 } - (\av{ A_{nn} })^2$ ``richtig'' ausgerechnet und 
%abgesch\"atzt werden kann? (siehe auch vorletzter Absatz).
%\PR{Finde ich weder im Kontext dieses Absatzes, noch des
%gesamten Papers wichtig genug. ETH ist ja an sich nicht wirklich unser 
%Thema hier.}]}

Next we return to the above-mentioned issues (i)-(iii).
The first of them regards the question of what can be
learned from the result for the ensemble average
$\av{\Abar}$ in (\ref{15})  with respect to the behavior of $\Abar$ for
the single members $V$ of the ensemble.
To this end, it is natural to consider the typical
deviations of the single $\Abar$'s from the average
behavior $\av{\Abar}$, which are, as usual, quantified by 
the variance
%$\av{(\Abar)^2}-\av{\Abar}^2$ 
$\av{(\,\Abar-\av{\Abar})^2}$,
or equivalently $\av{(\Abar)^2}-\av{\Abar}^2$.
Analogously to the previously discussed
mean value $\av{\Abar}$, the evaluation of the 
second moment $\av{(\Abar)^2}$ of the time-averaged 
expectation values from (\ref{5}) requires according 
to (\ref{6}) the ensemble averages over products of 
eight $U$-matrix elements,
and can again be achieved along the same 
general lines as in Ref. \cite{dab20}.
Referring once more to the Supplemental Material
of the latter paper for the detailed procedure,
one can show that
%Yet another important achievement of
%Ref. \cite{dab20} is the evaluation of the
%second moment $\av{(\Abar)^2}$
%of the time-averaged expectation values
%from (\ref{5}), which
%requires according to (\ref{6}) the ensemble 
%averages over products of eight $U$-matrix 
%elements. The main result of this calculation is 
%that the corresponding variance 
%$\av{(\Abar)^2}-\av{\Abar}^2$ 
the variance
%$\av{(\,\Abar-\av{\Abar})^2}$
is dominated by a factor of the form
$1/w$.
%, where $w$ is the interaction  strength from (\ref{9}).
%Disregarding extremely weak interactions in (\ref{3}),
%it thus 
With (\ref{10}) it
follows that when randomly sampling
a perturbation $V$ from the considered ensemble, 
then the time-averaged expectation value in (\ref{2}) 
is with very high probability very close to 
the ensemble averaged behavior from (\ref{15}).
In other words, the approximation 
\begin{eqnarray}
\Abar
        & \simeq & \av{\Abar}
%	\sum_{\mu\nu} \rho^0_{\mu\mu}(0)\, \tilde u(\mu-\nu)\, A^0_{\nu\nu} 
%	=\sum_{\nu} \tilde \rho^0_{\nu\nu}\, A^0_{\nu\nu} 
	\ ,
\label{17}
\end{eqnarray}
is very well satisfied for the vast majority of $V$'s.
This is our main result regarding the 
above-mentioned issue (i), see also the last paragraph of the paper.
%\PR{Altogether, the above-mentioned problem (i)
%is thus solved.}

As a next step, one can readily adapt the calculations from Ref. 
\cite{rei15} to show that, due to (\ref{10}),
%for sufficiently large interaction strengths $w$ in (\ref{9}), 
the vast majority 
of randomly sampled perturbations $V$ exhibit 
the following, so-called equilibration property:
After the initial transients have died out, the time-dependent
expectation value $\At$ from (\ref{1}) spends most of its 
time close to the time-averaged value $\Abar$
from (\ref{2}), symbolically indicated as
$\At\rightsquigarrow\Abar$ \cite{f4}.
It follows that the vast majority of perturbations even
simultaneously satisfies the approximations (\ref{17})
and $\At\rightsquigarrow\Abar$,
%and the approximation $\At\simeq\Abar$ for
%practically all sufficienty large times $t$,
symbolically summarized as
\begin{eqnarray}
\At\rightsquigarrow\av{\Abar}
\ \mbox{for most $V$'s.}
\label{18}
\end{eqnarray}
%Very roughly speaking, we thus got rid of the
%averages over time and over the $V$-ensemble
%in the original result by Deutsch as stated 
%below (\ref{16}).
In this sense, the above-mentioned problem 
(ii) can thus be considered as settled.

Turning to the remaining issue (iii) from above, this
problem would be solved immediately if the {\em unperturbed}
system were known to satisfy the ETH \cite{mor18,dal16,gog16}
(see also first paragraph).
%Finally, also the justification of the approximation
%(\ref{16}) in the text above Eq. (\ref{16})
%may be considered as not entirely satisfying:
%%from a somewhat more rigorous viewpoint:
%Even though the energy distribution 
%of $\tilde \rho$ is 
%%known to be quite 
%narrow and the level populations $\tilde \rho^0_{\nu\nu}$ 
%vary slowly as a function of $\nu$ 
%(see above Eq. (\ref{16})), the factors $A^0_{\nu\nu}$ 
%in (\ref{15}) may in principle still conspire in
%such a way that (\ref{16}) is violated.
%The problem is solved if the {\em unperturbed} 
%system is known to satisfy the ETH \cite{dal16,gog16,mor18} 
%(see also first paragraph). 
Namely, the ETH then guarantees that
the $A^0_{\nu\nu}$ are well approximated
by $\tr\{\rhomc A\}$ for all indices $\nu$ with 
non-negligible weights 
$\tilde\rho^0_{\nu\nu}$ in (\ref{15}).
Likewise, if the {\em perturbed} system is known 
to satisfy the ETH, then already in
Eq. (\ref{2}) the right-hand side will be
well approximated by the corresponding 
microcanonical expectation value.
More precisely speaking, all this 
regards the most common or
``strong'' version of the ETH (sETH)\cite{dal16,gog16,mor18}.
On the other hand, there also exists
a ``weak'' version of the ETH (wETH)
\cite{mor18,bri10,ike13,beu14,alb15,mor16,iko17,yos18,kuw20},
according to which the $A^0_{\nu\nu}$
may still notably deviate from $\tr\{\rhomc A\}$
for a small fraction of the summands
with non-negligible weights 
$\tilde\rho^0_{\nu\nu}$ in (\ref{15}).
But due to the fact that the 
$\tilde \rho^0_{\nu\nu}$ 
vary slowly as a function of 
$\nu$, this small fraction of
exceptional $A^0_{\nu\nu}$'s
in (\ref{15}) is not enough to
undermine the validity of (\ref{16}).

In other words,
it is sufficient that the unperturbed system satisfies the sETH or just the wETH
to guarantee (\ref{16}).
%However, until
Until now the sETH still has the
status of a hypothesis, which is supported 
by ample numerical evidence, but has not 
been proven for specific systems of actual 
interest \cite{dal16,gog16,mor18}.
In contrast, the wETH is known to 
be fulfilled at least for so-called translationally
invariant 
%spin-
Hamiltonians with 
short-range interactions 
%\cite{f5}
in combination with local observables
\cite{mor18,bri10,ike13,beu14,alb15,mor16,iko17,yos18,kuw20}.
In particular, both integrable and non-integrable
%such Hamiltonians $H_0$
Hamiltonians $H_0$ of this type
are thus still admitted
in (\ref{1}).
On the other hand, Hamiltonians $H_0$
%, while systems 
exhibiting many-body localization (MBL) are not
covered since they are not translationally 
invariant \cite{gog16,nan15}.
Altogether, issue (iii) is thus settled at least for 
those models which are known to obey the wETH.

On a final note,
we turn to the specific version of the
ETH as discussed by Deutsch \cite{deu91},
which actually amounts to yet another ``weak''
variant of the above-mentioned, most common 
sETH.
Namely, it is shown in Ref. \cite{deu91} that when
averaged over the $V$ ensemble, the perturbed diagonal matrix elements $A_{nn} := \bra{n} A \ket{n}$ of the observable $A$ are given by $\av{ \!A_{nn} } = \sum_\mu u(n-\mu) A^0_{\mu\mu}$ with $u(n)$ from~\eqref{8}.
Hence they exhibit a smooth dependence on the energy $E_n$ even if their unperturbed counterparts $A^0_{\mu\mu}$ vary considerably with $\mu$.
Moreover, by heuristic  arguments \cite{deu91} which can again be made rigorous along the lines of Ref.~\cite{dab20},
it can be shown that the corresponding variances $\av{A_{nn}^2} - \av{A_{nn}}^2$
are upper bounded by
$\Delta A^2 / \pi w$,
%{\bf [die Schranke, die ich habe, ist $\norm{A}^2 u(0)$; ich glaube, so, wie es vorher stand bzw. gemeint war, $\Delta A^2 / \pi w$, ist das Ergebnis, das man bei Deutsch findet; am Ende ist der Faktor $4$ auch egal ...]},
where $\Delta A$ denotes the spectral range of $A$, i.e., the difference between its largest and smallest eigenvalue.
In view of~\eqref{11},
this implies 
for an arbitrary but fixed $n$ that
\begin{equation}
\label{19}
	A_{nn} \simeq \sum_\mu u(n-\mu) \, A^0_{\mu\mu}
\end{equation}
for the vast majority of perturbed Hamiltonians $H$ from~\eqref{3}.
However, this result does not guarantee that the same
also applies {\em simultaneously} to sufficiently many $n$'s,
nor that the right-hand side in (\ref{19}) changes 
sufficiently slowly with $n$, as it is required by
the sETH.
As detailed in Ref.~\cite{rei15} this 
``weak'' variant of the ETH is thus not sufficient 
to imply thermalization for the vast majority 
of perturbed Hamiltonians $H$ in~\eqref{3}.
%As an aside it may also be mentioned that
%yet another ``weak'' variant of the ETH 
%was derived already in the paper by 
%Deutsch \cite{deu91} (see also first paragraph).
%An attempt to deduce thermalization directly 
%from this variant of the ETH was made 
%in Ref. \cite{rei15} but was not successful.

In conclusion, Deutsch's approach from Ref. \cite{deu91}
has been complemented so as to imply thermalization
for a very large class of perturbed 
Hamiltonians (\ref{3}), initial 
conditions $\rho(0)$, and 
observables $A$.
In particular, with respect to the perturbations $V$,
a large variety of random matrix ensembles were 
admitted.
As usual in random matrix theory \cite{bor16}, it is thus
reasonable to expect that the same behavior
will also be recovered in some given
(non-random)
%specific 
model Hamiltonians of the form (\ref{3}).
Providing a more rigorous justification of this well-established
common lore in random matrix theory is a long-standing, 
very difficult task, and is beyond the scope of our present
work and of Deutsch's original 
approach \cite{deu91}.
On the other hand, it may be worthwhile 
to point out some obvious circumstances under 
%which a failure of our present general framework is expected:
which such a line of reasoning is bound to fail:
Namely, when the given model Hamiltonian 
in (\ref{3}) exhibits certain ``special features'' 
(symmetries, constants of motion, etc.)
which are known to generically rule out 
thermalization, while the same features 
are no longer exhibited by most of the
random perturbations $V$.
In particular, this applies to models which are 
integrable \cite{mor18,dal16} or exhibit many-body 
quantum scars \cite{khe20}.

%%%%%%%%%%%%%%%%%%%%%%%%%%%%%%%%%%%%%%%%%%%%%%%%%%%%

\begin{acknowledgments}

This work was supported by the 
Deutsche Forschungsgemeinschaft (DFG)
within the Research Unit FOR 2692
under Grant No. 397303734
%, by the Paderborn Center for Parallel 
%Computing (PC$^2$) within the Project 
%HPC-PRF-UBI2,
and by the International Centre for Theoretical Sciences 
(ICTS) during a visit for the program -  
Thermalization, Many body localization and Hydrodynamics 
(Code: ICTS/hydrodynamics2019/11).

\end{acknowledgments}

%%%%%%%%%%%%%%%%%%%%%%%%%%%%%%%%%%%%%%%%%%%%%%%%%%%%
%%%%%%%%%%%%%%%%%%%%%%%%%%%%%%%%%%%%%%%%%%%%%%%%%%%%


\begin{thebibliography}{99}


\bibitem{tas16}
H. Tasaki,
Typicality of Thermal Equilibrium and Thermalization in Isolated 
Macroscopic Quantum Systems,
J. Stat. Phys. {\bf 163}, 937 (2016).

\bibitem{mor18}
T. Mori, T. N. Ikeda, E. Kaminishi, and M. Ueda,
Thermalization and prethermalization 
in isolated quantum systems: a theoretical overview,
J. Phys. B  {\bf 51}, 112001 (2018).

\bibitem{dal16}
L. D'Alessio, Y. Kafri, A. Polkovnikov, and M. Rigol,
From Quantum Chaos and Eigenstate Thermalization
to Statistical Mechanics and Thermodynamics,
Adv. Phys.  {\bf 65}, 239 (2016).

\bibitem{gog16}
C. Gogolin and J. Eisert,
Equilibration, thermalization, and the emergence
of statistical mechanics in closed quantum systems,
Rep. Prog. Phys. {\bf 79}, 056001 (2016).

\bibitem{bor16}
F. Borgonovi, F. M. Izrailev, L. F. Santos, and V. G. Zelevinsky,
Quantum chaos and thermalization in isolated systems of 
interacting particles,
Phys. Rep.  {\bf 626}, 1 (2016).

\bibitem{deu91}
J.~M.~Deutsch, 
Quantum statistical mechanics in a closed system,
Phys. Rev. A {\bf 43}, 2046 (1991); 
J.~M.~Deutsch, 
A closed quantum system giving ergodicity,
deutsch.physics.ucsc.edu/pdf/quantumstat.pdf;
%%\verb+physics.ucsc.edu/~josh/pdf/quantumstat.pdf+
%%physics.ucsc.edu/\verb+~+josh/pdf/quantumstat.pdf
%(unpublished);
J.~M.~Deutsch,  
Thermodynamic entropy of a many-body 
energy eigenstate, New J. Phys. {\bf 12}, 075021 (2010).

\bibitem{f0}
Numerous specific examples for the closed 
many-body  quantum systems one usually 
has in mind in this context can be found, 
for instance, in the comprehensive Reviews
\cite{tas16,mor18,dal16,gog16,bor16}.
Particular examples adopting the present random matrix 
ensembles to emulate concrete physical systems are available, 
among others, in Ref. \cite{dab20}.

\bibitem{f1}
Implicitly, the labels have been chosen so that
the $E_n$'s are ordered by magnitude and approach
$E^0_{\nu=n}$ for $V\to 0$ in (\ref{3}).

\bibitem{f2}
Some factors of $2$ are missing in the 1991 paper,
but not any more in the 2010 paper from Ref. \cite{deu91}.

\bibitem{dab20}
L. Dabelow and P. Reimann,
Relaxation Theory for Perturbed Many-Body Quantum Systems versus Numerics and Experiment,
Phys. Rev. Lett. {\bf 124}, 120602 (2020).

\bibitem{wig55}
E. P. Wigner,
Characteristic vectors of bordered matrices 
with infinite dimensions,
Ann. Math.  {\bf 62}, 548 (1955).

\bibitem{fyo96}
Y. V. Fyodorov, O. A. Chubykalo, F. M. Izrailev, and G. Casati,
Wigner random banded matrices with sparse structure: 
local spectral density of states,
Phys. Rev. Lett.  {\bf 76}, 1603 (1996).

\bibitem{fyo95}
%\bibitem{she94}
%D. L. Shepeliansky,
%{ Coherent propagation of two interacting particles in a random potential},
%{\em Phys. Rev. Lett. {\bf 73}, 2607 (1994);
Y. V. Fyodorov and A. D. Mirlin,
Statistical properties of random banded 
matrices with strongly fluctuating diagonal elements,
Phys. Rev. B  {\bf 52}, R11580 (1995).

%\bibitem{f3}
%\PR{More precisely speaking, 
%it was pointed out already in Refs. \cite{wig55,deu91}
%and elaborated more rigorously in
%Ref. \cite{fyo96} that a finite band width 
%%inevitably must give 
%gives
%rise to a crossover
%of the relatively slow decay of (\ref{8}) 
%with $n$ into a much faster (essentially 
%exponential) decay for sufficiently 
%large $n$, and similarly for
%$\tilde u(n)$ in (\ref{12}).
%Since it is reasonable to expect that
%this scenario applies very generally
%to physically realistic models \cite{deu91},
%possible concerns regarding the slowly 
%decaying tails of $\tilde u(n)$ in 
%Eqs. (\ref{11}), (\ref{14}), and (\ref{15})
%can thus be readily 
%overcome.}

\bibitem{nat18}
C. Nation and D. Porras,
Off-diagonal observable elements from random 
matrix theory: distributions, fluctuations, and 
eigenstate thermalization,
New. J. Phys.  {\bf 20}, 103003 (2018).

%\bibitem{ith18}
%\LD{G. Ithier and S. Ascroft,
%Statistical diagonalization of a random biased Hamiltonian: the case of the eigenvectors,
%J. Phys. A: Math. Theo. {\bf 51}, 48LT01 (2018).}

\bibitem{rei15}
P. Reimann,
Eigenstate thermalization: Deutsch's approach and beyond,
New J. Phys.  {\bf 17}, 055025 (2015).

\bibitem{f4}
Note that a small fraction of exceptional times $t$ is 
unavoidable, e.g., due to quantum revivals, caused 
by the quasiperiodicity of $\At$ in (\ref{1}).

%\bibitem{f5}
%More precisely, for spin-systems in more than
%one dimension, one must also require \cite{mor16}
%that  the energy not too small (excluding 
%spontaneous symmetry breaking induced 
%phase transitions in the thermodynamic limit).

\bibitem{bri10}
G. Biroli, C. Kollath, and A. M. L\"auchli, 
Effect of Rare Fluctuations on the Thermalization of Isolated Quantum Systems, 
Phys. Rev. Lett. {\bf 105}, 250401 (2010).

\bibitem{ike13}
T. N. Ikeda, Y. Watanabe, and M. Ueda, 
Finite-size scaling analysis of the eigenstate thermalization 
hypothesis in a one-dimensional interacting Bose gas, 
Phys. Rev. E {\bf 87}, 012125 (2013).

\bibitem{beu14}
W. Beugeling, R. Moessner, and M. Haque, 
Finite-size scaling of eigenstate thermalization, 
Phys. Rev. E {\bf 89}, 042112 (2014).

\bibitem{alb15}
V. Alba, 
Eigenstate thermalization hypothesis and integrability in quantum spin chains, 
Phys. Rev. B {\bf 91}, 155123 (2015).

\bibitem{mor16}
T. Mori,
Weak eigenstate thermalization with large deviation bound,
arXiv:1609.09776

\bibitem{iko17}
E. Iyoda, K. Kaneko, and T. Sagawa,
Fluctuation Theorem for Many-Body Pure Quantum States,
Phys. Rev. Lett. {\bf 119}, 100601 (2017).

\bibitem{yos18}
T. Yoshizawa, E. Iyoda, and T. Sagawa,
Numerical large deviation analysis of eigenstate 
thermalization hypothesis,
Phys. Rev. Lett {\bf 120}, 200604 (2018).

\bibitem{kuw20}
T. Kuwahara and K. Saito,
Eigenstate thermalization from clustering property of correlations,
Phys. Rev. Lett. {\bf 124}, 200604 (2020).

\bibitem{nan15}
R. Nandkishore and D. A. Huse,
Many-body localization and thermalization 
in quantum statistical mechanics,
Annu. Rev. Condens. Matter Phys. {\bf 6}, 15 (2015).

\bibitem{khe20}
V. Khemani, M. Hermele, and R. Nandkishore,
Localization from Hilbert space shattering: From theory to physical realizations,
Phys. Rev. B {\bf 101}, 174204 (2020).

\end{thebibliography}
\end{document}